\documentclass[aps,prb,twocolumn,amssymb,superscriptaddress,floatfix,a4paper,showpacs]{revtex4-1}

\usepackage{graphicx}
\usepackage{bm}
\usepackage{amsmath}

\usepackage{color}

\begin{document}

\title{Thermal rounding of the depinning transition in ultrathin Pt/Co/Pt films}
\author{S. Bustingorry}
\affiliation{CONICET, Centro At{\'{o}}mico Bariloche, 8400 San Carlos de Bariloche, R\'{\i}o Negro, Argentina}
\author{A. B. Kolton}
\affiliation{CONICET, Centro At{\'{o}}mico Bariloche, 8400 San Carlos de Bariloche, R\'{\i}o Negro, Argentina}
\author{T. Giamarchi}
\affiliation{DPMC-MaNEP, University of Geneva, 24 Quai Ernest Ansermet, 1211 Geneva 4, Switzerland}

\date{\today}

\begin{abstract}
We perform a scaling analysis of the mean velocity of extended magnetic domain walls driven in ultrathin Pt/Co/Pt ferromagnetic films with perpendicular anisotropy, as a function of the applied external field for different film-thicknesses. We find that the scaling of the experimental data around the thermally rounded depinning transition is consistent with the universal depinning exponents theoretically expected for elastic interfaces described by the one-dimensional quenched Edwards-Wilkinson equation. In particular, values for the depinning exponent $\beta$ and thermal rounding exponent $\psi$ are tested and the present analysis of the experimental data is compatible with $\beta=0.33$ and $\psi=0.2$, in agreement with numerical simulations.

\end{abstract}


\maketitle

% \tableofcontents

\section{Introduction}\label{sec:intro}
Elastic manifolds weakly pinned by random impurities are ubiquitous in condensed matter physics. Paradigmatic examples are ferromagnetic and ferroelectric domain walls,~\cite{lemerle_domainwall_creep,bauer_deroughening_magnetic2,yamanouchi_creep_ferromagnetic_semiconductor2,metaxas_depinning_thermal_rounding,metaxas_coupled_interfaces,paruch_ferro_roughness_dipolar,paruch_ferro_quench,guyonnet_domainwall_ferro_shear,guyonnet_domainwall_ferro} superconducting vortex lattices,~\cite{blatter_vortex_review,giamarchi_vortex_review,du_aging_bragg_glass} charge density waves,~\cite{nattermann_cdw_review} contact lines of liquid menisci,~\cite{moulinet_distribution_width_contact_line2,ledoussal_contact_line} and fractures.~\cite{bouchaud_crack_propagation2,ponson_fracture,alava_review_cracks} No matter how weak disorder is, in all these systems the key factor determining their dynamical properties is the competition between elasticity, which tends ``to align'' the displacement field into a perfectly flat or periodic structure, and disorder, which tends to distort it. This gives rise to universal glassy properties with characteristic rough structures and complex collective pinning phenomena.

One of the most remarkable and possibly the most accurate experimental verification for the universal glassy dynamics theoretically predicted for disordered elastic systems has been the measurement of the ultra-slow {\it creep motion} of magnetic domain walls in ultrathin Pt/Co/Pt ferromagnetic films with perpendicular anisotropy driven by a very small (well below the depinning threshold defined below) applied magnetic fields.~\cite{lemerle_domainwall_creep} The so-called creep law states that the mean velocity $V$ of an extended elastic interface in a random environment follows a strongly non-linear function of the field
\begin{equation}
 V \sim \exp \left[ -\frac{U_c}{k_B T} \left( \frac{H_c}{H} \right)^\mu
\right],
\label{eq:creep_law}
\end{equation}
where $U_c$ gives a characteristic energy scale in the creep regime and the applied field is $H \ll H_c$, with $H_c$ the so-called depinning threshold. Both $U_c$ and $H_c$ are material dependent parameters and increase with the strength of the disorder. Physically, this law can be interpreted as an Arrhenius activated motion over typical energy barriers separating different metastable states. The particular form of the creep law is directly related to the divergence of these energy barriers with decreasing applied field $H$.~\cite{feigelman_collective,nattermann_creep_law,narayan_fisher_cdw,nattermann_stepanow_depinning,chauve_creep_short,chauve_creep_long,kolton_creep2} This law is universal in the sense that the exponent $\mu$ only depends on the dimensionality $d$ of the elastic manifold and its equilibrium roughness exponent $\zeta_{eq}$ through the relation
\begin{equation}
 \mu = \frac{d-2+2\zeta_{eq}}{2-\zeta_{eq}}
\label{eq:creep_law_exponent}
\end{equation}
The roughness exponent $\zeta_{eq}$ measures the rate at which the interface width grows with its linear size $w\sim L^{\zeta_{eq}}$ at equilibrium and is in turn universal: it depends only on $d$, on the nature of the disorder correlations and on the short-ranged character of elastic interactions. This makes this experimental system a paradigmatic example of the universal physics predicted for elastic manifolds weakly pinned by random impurities.

The theory for driven disordered elastic systems applied to magnetic domain walls also predicts two additional dynamical regimes as a function of $H$.~\cite{agoritsas_review} For large fields $H \gg H_c$, in the so-called {\it fast-flow} regime, the disorder acts effectively as a fictive thermal noise and the response is linear
\begin{equation}
V \approx m H,
\end{equation}
with $m$ the domain wall mobility characterizing dissipation processes during the flow motion. The nature of the involved dissipation processes in the fast-flow regime depends on the value of the Walker field $H_W$,~\cite{slonczewski_walker_field,schryer_walker_field} separating steady flow for $H<H_W$ and precessional flow for $H>H_W$. The analysis of the fast-flow regime in ultrathin Pt/Co/Pt ferromagnetic films suggests that $H_W \ll H_c$, which means that one only has access to the precessional flow at large fields and the steady flow can not be observed because it is well inside the creep regime.~\cite{metaxas_depinning_thermal_rounding}

Finally, around the critical field $H_c$ a non-trivial zero-temperature {\it depinning regime} is expected. In this regime the velocity follows a power-law behavior when approaching $H_c$ from above, $V \sim (H-H_c)^\beta$ with $\beta$ an universal critical exponent, analogous to continuous phase transitions.~\cite{fisher_depinning_meanfield} However, unlike creep motion, the latter scaling was not experimentally tested yet, and to the best of our knowledge, the data set reported in Ref.~\onlinecite{metaxas_depinning_thermal_rounding} give the first experimental evidence supporting $\beta<1$, as predicted by theory.~\cite{narayan_fisher_cdw,nattermann_stepanow_depinning,chauve_creep_long} Since $\beta$ is essentially a zero-temperature quantity it has not been \textit{a priori} obvious how to obtain it in experiments. This leads us to the fundamental issue of the temperature effects on the depinning transition. The naive analogy with standard phase transitions (eg. magnetization vs temperature with an external magnetic field in the Ising model), thinking $V$ as the order parameter and $H$ as the control parameter suggests that $V \sim T^\psi$ at $H=H_c$, with $\psi$ an universal thermal rounding exponent.~\cite{middleton_CDW_thermal_exponent,chen_marchetti,nowak_thermal_rounding,roters_thermal_rounding1,vandembroucq_thermal_rounding_extremal_model,luo_thermal_rounding_flux_lines,bustingorry_thermal_rounding_epl,bustingorry_thermal_depinning_exponent} Although the very existence and universality of such power-law has not been rigorously proven for the depinning transition, it is consistent with recent numerical simulations.~\cite{roters_thermal_rounding1,luo_thermal_rounding_flux_lines,bustingorry_thermal_rounding_epl,bustingorry_thermal_depinning_exponent}

Furthermore, based on this analogy with standard phase transitions one can argue that the order parameter is an homogeneous function of the external parameters. This leads to universal scaling functionality for the order parameter around the critical point. In the present case one generally arrives at the thermal rounding scaling form
\begin{equation}
V \sim T^{\psi} G \left( \frac{H-H_c}{T^{\psi/\beta}} \right),
\label{eq:thermal_rounding_scaling}
\end{equation}
with $G(x) \sim \mathrm{const.}$ for small $x$ and $G(x)\sim x^\beta$ for large $x$. This scaling behavior is expected to hold close to the critical point, i.e. for  $H-H_c \ll H_c$ and $T \to 0$. Although this scaling form has been successfully tested in numerical simulations,~\cite{roters_thermal_rounding1,luo_thermal_rounding_flux_lines,bustingorry_thermal_depinning_exponent} it has not yet been experimentally probed.

An experimental test of Eq.~\eqref{eq:thermal_rounding_scaling} for a system in which the creep-law can be accurately verified is important for several reasons. First, it allows to identify more precisely the non-equilibrium universality class and to further explore the regime of applicability of the elastic theory which provides precise enough predictions for the exponents $\beta$ and $\psi$. Indeed, although creep experiments of Refs~\onlinecite{lemerle_domainwall_creep,metaxas_depinning_thermal_rounding} are consistent with the universality class of one-dimensional elastic interfaces with short-ranged elasticity and short-ranged random-bond disorder, this is not yet enough to determine the depinning universality class. This can also depend on the precise nature of elastic interactions and/or the presence of additional terms in the equation of motion which might be irrelevant at the length-scales probed by transport in the creep regime or at the $H=0$ equilibrium regime. Second, recent theoretical predictions show that equilibrium, depinning, and fast-flow type of motions can not be separated below the depinning threshold, but they manifest at different velocity-dependent characteristic length-scales, thus breaking the quasi-equilibrium picture of creep motion.~\cite{kolton_depinning_zerot2,kolton_dep_zeroT_long} Although the leading dominant part of the creep law is not expected to change by these new predictions, the large scale geometry of the moving interface is expected to display the same depinning roughness exponents that describe it around $H_c$.~\cite{kolton_depinning_zerot2,kolton_dep_zeroT_long} Finally, and specially for one-dimensional interfaces, the validity of the elastic approximation, which underlies the prediction of Eqs.\eqref{eq:creep_law} and \eqref{eq:creep_law_exponent}, is not experimentally evident for the relevant length-scales tested around depinning, where interfaces may become boundlessly rough, as found in some minimal models (eg. quenched Edwards-Wilkinson equation~\cite{edwards_wilkinson,Barabasi-Stanley}).

We report here a scaling analysis of the experimental data reported in Refs.~\onlinecite{metaxas_depinning_thermal_rounding,metaxas_thesis} for the mean velocity of extended magnetic domain walls driven in ultrathin Pt/Co/Pt ferromagnetic films with perpendicular anisotropy, as a function of the applied external field, for different film-thicknesses. We find that the data around the thermally rounded depinning transition is consistent with the universal depinning exponents $\beta$ and $\psi$ theoretically expected for elastic interfaces described by the one-dimensional quenched Edwards-Wilkinson equation~\cite{duemmer2,bustingorry_thermal_rounding_epl,bustingorry_thermal_depinning_exponent}. The manuscript is organized as follows. First, in Sec.~\ref{sec:expdata} we present the experimental data obtained in Ref.~\onlinecite{metaxas_depinning_thermal_rounding} together with the physical parameters of the system that we will use in the scaling analysis. Then in Sec.~\ref{sec:new-values} we present new estimated values for the critical depinning field based on the scaling analysis of the experimental data together with a fitted value for the thermal rounding exponent. In Sec.~\ref{sec:scal} we show that the scaling relation Eq.~\eqref{eq:thermal_rounding_scaling} satisfactorily describes the experimental data around depinning. Finally, Sec.~\ref{sec:conc} is devoted to conclusions and final comments.

\section{Experimental data and key quantities}
\label{sec:expdata}

\begin{figure}[!tbp]
\includegraphics[width=7.5cm,clip=true]{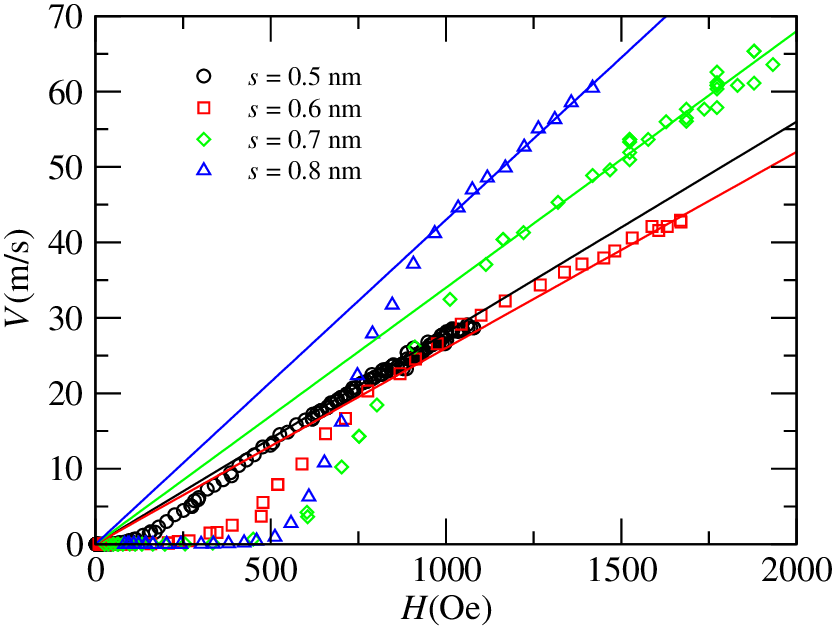}
\caption{ (color online) Experimental velocity-fields curves for driven domain wall motion in ultrathin Pt/Co/Pt films with perpendicular anisotropy. Each curve corresponds to different thickness for the Co layer, as indicated. Continuous lines represent linear fits to the fast-flow regime at large fields. This experimental data had been reported in Ref.~\onlinecite{metaxas_depinning_thermal_rounding}.}
\label{fig:exp-data}
\end{figure}

Briefly, the velocity-field curves were obtained at room temperature by imaging magnetic domains using a high-resolution far-field PMOKE microscope with a CCD camera before and after the application of field pulses of equal magnitude and different duration. The images were subtracted to measure the displacements of the domain wall and the velocity were obtained from this displacement and the field pulse duration. Further details of the experimental set-up can be found in Refs.~\onlinecite{metaxas_depinning_thermal_rounding,metaxas_thesis}.

Figure~\ref{fig:exp-data} shows the experimental results for the velocity-force characteristics of ultrathin Pt/Co/Pt ferromagnetic films.~\cite{metaxas_depinning_thermal_rounding} Different curves correspond to different thickness of the Co layer, $s=0.5,0.6,0.7$ and $0.8$nm. One can observe different driven regimes on all curves. In particular the creep, depinning and fast-flow regimes can be clearly identified on each curve. The depinning regime becomes smeared and not abrupt due to thermal activation; this is the so-called thermally rounded depinning transition. Roughly, the sharp increase of the velocity can be interpreted as the crossover from the thermally activated creep regime to the thermally rounded depinning transition. The beginning of this crossover can be identified by the depart from the creep-law Eq.~\eqref{eq:creep_law}, only valid for very small fields.~\cite{metaxas_depinning_thermal_rounding} Interestingly, for larger fields, the curvature of the $V(H)$ curves changes from positive to negative before crossing over to the fast-flow linear regime, resembling the thermally rounded depinning transition with an exponent $\beta<1$ observed in simple elastic models. All these features support our attempt to fit this data with the tools used in standard phase transitions as summarized by Eq.~\eqref{eq:thermal_rounding_scaling}.

The main difficulty to attempt a quantitative analysis of the thermally rounded depinning transition from the available data for different samples of the same material is to obtain proper rescaled variables. We need relatively precise values for the material dependent quantities such as the characteristic temperature $T_c$, the domain wall mobility $m$ and in particular the critical field $H_c$ which might all depend on the film thickness $s$. To this end we first analyze the fast-flow regime and the creep regime in order to get reasonable bounds for the key parameters.

From the fast-flow regime, where the velocity is proportional to the force, one can get the mobility $m$, by fitting $V = m H$ to the data. This has been already reported in Ref.~\onlinecite{metaxas_depinning_thermal_rounding}. As we can appreciate in Fig.~\ref{fig:exp-data}, $m$ seems to increase when increasing the sample thickness $s$. Values for the mobility $m$ are shown in Table~\ref{tab:exp} for future reference on this work.

One can also observe in Fig.~\ref{fig:exp-data} that the sudden increase in the velocity-field curves beyond the creep regime is reached at different field values. This means that each $V(H)$ curve has a thickness dependent critical depinning field $H_c$. The proper estimation of $H_c$ at finite temperatures is a difficult task. Indeed, in Ref.~\onlinecite{metaxas_depinning_thermal_rounding} the authors reported, as a lower bound for $H_c$, the characteristic value $H^*$, where the non-linear fit of the creep-law starts failing. Values for $H^*$, extracted from Ref.~\onlinecite{metaxas_depinning_thermal_rounding}, are reproduced in Table~\ref{tab:exp} for reference. As one can appreciate the general trend is that $H^*$ increases with increasing $s$.

\begin{table}
\caption{\label{tab:exp} For each film thickness $s$, we present here the key experimental parameters needed to perform the scaling analysis of the bare velocity-filed curves shown in Fig.~\ref{fig:exp-data}. $m$ is the mobility of the domain wall in the fast-flow regime, $H^*$ is a lower bound for the critical depinning field, and $T_c/T$ is the effective inverse temperature; all this values were reported in Ref.~\onlinecite{metaxas_depinning_thermal_rounding}. Values for the critical depinning fields $H_c$ and the effective inverse temperature $T_c/T$ as obtained with the scaling approach used in this work are given. We also include values for the phenomenological critical field $H_c^{\mathrm{phen}}$ which compares well with $H_c$.}
\begin{ruledtabular}
\begin{tabular}{cccccc}
$s(\mathrm{nm})$&0.5&0.6&0.7&0.8&\\
\hline
$m(\mathrm{m/s Oe})$&0.028&0.026&0.034&0.043&Ref.~\onlinecite{metaxas_depinning_thermal_rounding}\\
$H^*(\mathrm{Oe})$&230&590&750&650&Ref.~\onlinecite{metaxas_depinning_thermal_rounding}\\
$T_c/T$&9&14&22&35&Ref.~\onlinecite{metaxas_depinning_thermal_rounding}\\
\hline
$H_c(\mathrm{Oe})$&330&660&800&730&this work\\
$T_c/T$&8.2&13.6&21.6&34.0&this work\\
$H_c^{\mathrm{phen}}(\mathrm{Oe})$&320&670&840&730&this work\\
%In& 0.460 & 18.40 & 3.500 &Ba\footnotemark[5]
\end{tabular}
\end{ruledtabular}
%\footnotetext[5]{And etc.}
\end{table}

Furthermore, as shown in Ref.~\onlinecite{metaxas_depinning_thermal_rounding}, in the small-field creep regime the velocity is well fitted by the one dimensional case of Eq.~\eqref{eq:creep_law_exponent}, with $\mu=1/4$ for $d=1$, as expected for the well known one-dimensional equilibrium roughness exponent $\zeta_{eq}=2/3$. Therefore, if the value of $H_c$ is known, then one can use the creep formula Eq.~\eqref{eq:creep_law} to fit the value of $T_c/T$, with $T_c=U_c/k_B$ the characteristic temperature scale associated to the typical energy scale $U_c$.~\cite{feigelman_collective,nattermann_creep_law} Previously, the lower bound $H^*$ (instead of the depinning field $H_c$) had been used to fit $T_c/T$ using the creep formula.~\cite{metaxas_depinning_thermal_rounding} This is indeed a good approximation since given that $H^* \lesssim H_c$ and $\mu$ is small, $T_c/T$ can only be corrected by a small factor of order $(H^*/H_c)^\mu$. The values for $T_c/T$ obtained using $H^*$ are shown in Table~\ref{tab:exp}. As one can observe, $T_c/T$ increases with increasing $s$. This is indicating that although all curves were obtained at room temperature, the effective temperature $T/T_c$ is decreasing with the film thickness $s$ due to a change in the intrinsic disorder energy scale. In consequence, each curve can be thought as being at a different temperature, allowing to effectively test the thermally rounded depinning regime.

In this work we need to use the experimental values shown in Table~\ref{tab:exp} for the scaling analysis of the data in Fig.~\ref{fig:exp-data}. However, since the thermal rounding scaling of the depinning transition critically depends on the value of $H_c$ we will not use the previously reported value $H^*$, which actually corresponds to a lower bound for $H_c$. Instead, we will use a different approach trying to get a better estimate for the depinning field $H_c$. This new value will also permit to estimate the thermal rounding exponent $\psi$ as shown in the following Section.

\section{Experimental estimates for $H_c$ and $\psi$}
\label{sec:new-values}

In the literature, mainly three ways of estimating the value of the critical field $H_c$ from $V(H)$ curves were used: (i) a linear extrapolation to zero velocity, $V \propto (H-H_c)$, from data around depinning,~\cite{lemerle_domainwall_creep} (ii) when choosing a critical velocity $V_c$, one can determine at low temperatures the critical field from $V(H_c)=V_c$,~\cite{kirilyuk_magnetization_reversal,krusin_pinning_wall_magnet} and (iii) the end of the creep regime from below as a lower bound for $H_c$.~\cite{metaxas_depinning_thermal_rounding} The first two protocols are useful when one has velocity-field curves over a limited field range. Here, in order to obtain $H_c$ from the experimental data shown in Fig.~\ref{fig:exp-data}, we will adopt a different approach, which is based on scaling concepts and relies on the knowledge of a precise value for the depinning exponent $\beta$. Then we will show that this is also compatible with a simple phenomenological determination of the depinning field.

\begin{figure}[!tbp]
\includegraphics[width=7.5cm,clip=true]{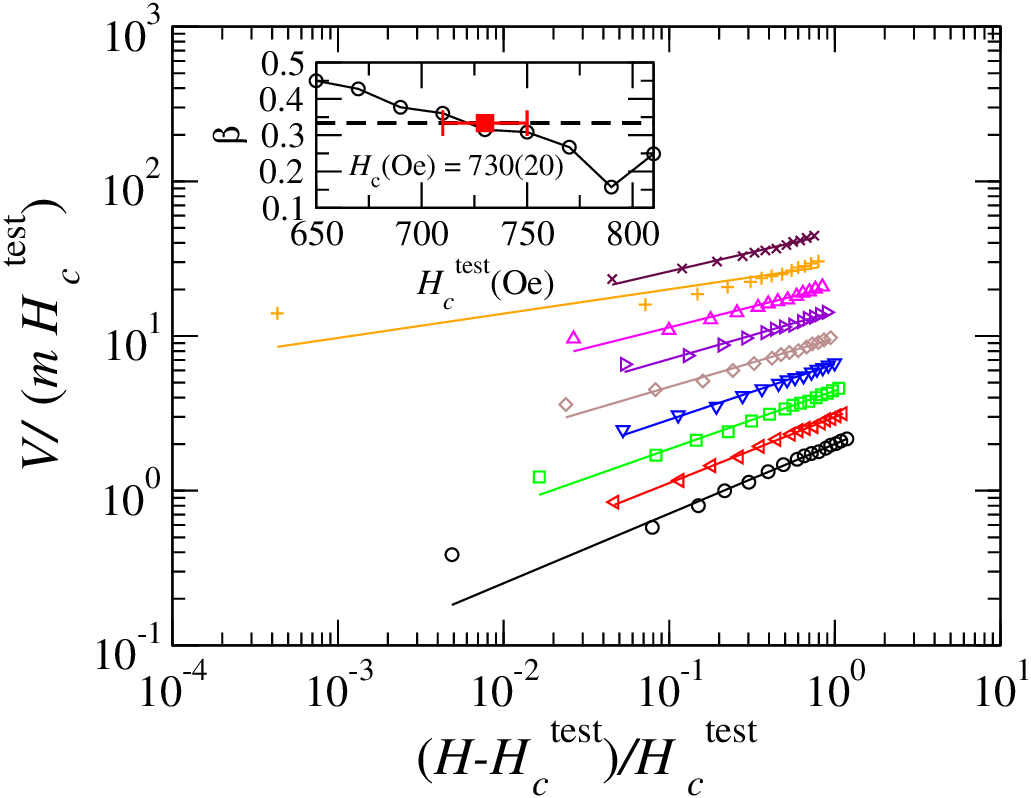}
\caption{ (color online) Fitting the value of the depinning field $H_c$ for the sample with $s=0.8$ while searching the best fit to the theoretically expected value for the depinning exponent $\beta \approx 1/3$. The main panel shows fitting curves for different test depinning fields in scaled variables, with $650 \mathrm{Oe} < H_c^{\mathrm{test}} < 810 \mathrm{Oe}$ from bottom to top in steps of $20 \mathrm{Oe}$ (data were shifted upwards for clarity). The inset presents the obtained values for $\beta(H_c^{\mathrm{test}})$. The red square point in the inset corresponds to the estimated depinning field $H_c$ for $s=0.8$nm.}
\label{fig:hc}
\end{figure}

Although the value $\mu=1/4$ for the creep regime is widely supported by experimental results,~\cite{lemerle_domainwall_creep,metaxas_depinning_thermal_rounding} an experimental estimated value for the depinning exponent $\beta$ has not been reported. In fact, to the best of our knowledge, the data in Fig.~\ref{fig:exp-data} give the first experimental support to $\beta<1$ as theoretically expected in low dimensional systems.~\cite{chauve_creep_long} In spite of the fact that $\beta$ is strictly a zero temperature quantity one can still obtain it by fitting the power-law behavior, $V \sim (H-H_c)^\beta$ at very small temperatures $T \ll T_c$. As noted in the values reported in Table~\ref{tab:exp}, the effective reduced temperature scales are always relatively small, $T/T_c < 0.13$. If one takes the lower bound $H^*$ and try to fit $\beta$ from $V \propto (H-H^*)^\beta$ the obtained value is systematically larger than the one dimensional accepted value $\beta = 0.33$, which is based on numerical simulations of the quenched Edwards-Wilkinson equation.~\cite{duemmer2} The obtained values, using $H^*$ are $\beta^*=0.58$, $0.44$, $0.35$ and $0.45$ for $s=0.5$, $0.6,$ $0.7$ and $0.8$nm respectively. Only the value for $s=0.7$ is close to $\beta \approx 1/3$.

Here, in order to obtain $H_c$, we assume deviations of the experimental domain wall from the elastic model (such as overhangs, bubbles, etc.) do not play an important role. We also assume that magnetic domain walls, as smooth elastic objects with short-range elasticity and uncorrelated disorder, are well described by the one-dimensional quenched Edwards-Wilkinson equation. Indeed, this model successfully reproduce the creep law observed in this system. With these two assumptions, whose validity will be checked by they consistency, we take the following protocol: we search for the best value of $H_c$ which permits to fit $\beta$ closer to $1/3$. This protocol is illustrated in Fig.~\ref{fig:hc} for the case $s=0.8$, whose bare velocity-field curve is shown in Fig.~\ref{fig:exp-data}. For the fitting procedure, we first discard the large-field data points, $H > 1000$Oe, corresponding to the fast-flow regime. Then we propose as a first approximate depinning field $H_c^{\mathrm{test}}=H^*$, and we plot the remaining data for $H > H_c^{\mathrm{test}}$ using scaled variables, $V/(m H_c^{\mathrm{test}})$ against $(H-H_c^{\mathrm{test}})/H_c^{\mathrm{test}}$, which corresponds to the lower curve of Fig.~\ref{fig:hc}. From this curve we fit the $\beta(H_c^{\mathrm{test}})$ value. Repeating this for increasing values of $H_c^{\mathrm{test}}$ we obtain a set of points $\beta(H_c^{\mathrm{test}})$ as shown in the inset of Fig.~\ref{fig:hc}. $\beta(H_c^{\mathrm{test}})$ is decreasing with $H_c^{\mathrm{test}}$ and intersects the theoretically expected value for $\beta$. Therefore, from this curve we extract the value of the critical field $H_c$ as that which corresponds to $\beta$ closer to $1/3$. In this case we estimate for $s=0.8$ the value $H_c=730(20)$Oe, as shown by the red square point in the inset of Fig.~\ref{fig:hc}.

\begin{figure}[!tbp]
\includegraphics[width=7.5cm,clip=true]{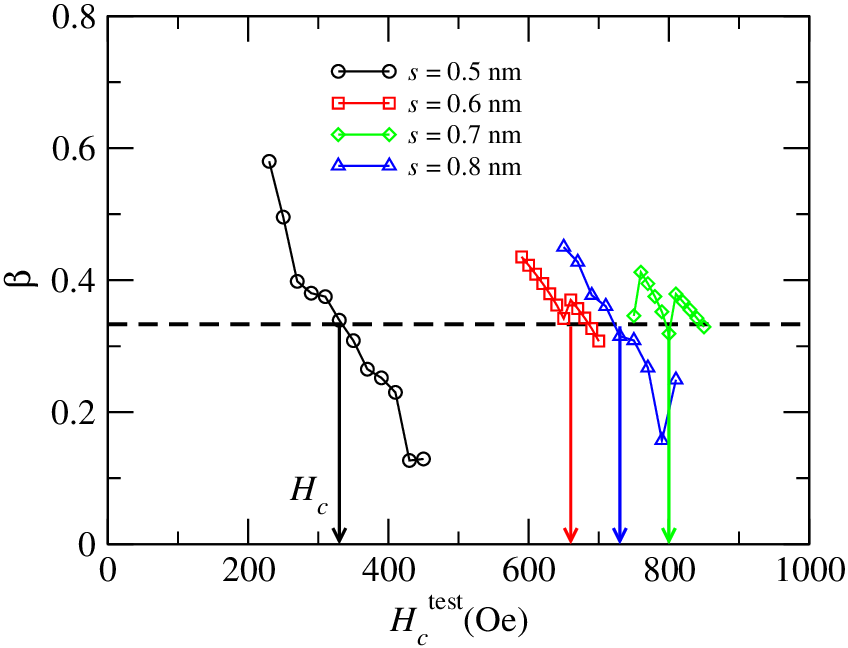}
\caption{ (color online) Dependence of the depinning exponent $\beta$ on the test field $H_c^{\mathrm{test}}$ for different film thickness as indicated. This permits to estimate the new values for the critical fields $H_c$ (indicated by the vertical arrows) as the values when $\beta$ is closer to the $\beta = 1/3$ value (dashed line). The lowest field point of each curve corresponds to the lower bound for the depinning field $H_c^{\mathrm{test}}=H^*$.}
\label{fig:beta-hc}
\end{figure}

Following this protocol, we present in Fig.~\ref{fig:beta-hc} the obtained $\beta(H_c^{\mathrm{test}})$ data for different thickness values as indicated. The lowest field point of each curve corresponds to the lower bounds for the depinning field $H_c^{\mathrm{test}}=H^*$. The vertical arrows indicate the estimated values $H_c$ for different $s$ values, which are given in Table~\ref{tab:exp}. We have also tested that this protocol gives a good value for the critical field in numerical simulations, where a precise value of the critical field can be obtained by exact algorithms. Finally, we also show in Table~\ref{tab:exp} slightly corrected values for the effective inverse temperature scale $T_c/T$ obtained using the new estimate for the critical field, $H_c$.

\begin{figure}[!tbp]
\includegraphics[width=7.5cm,clip=true]{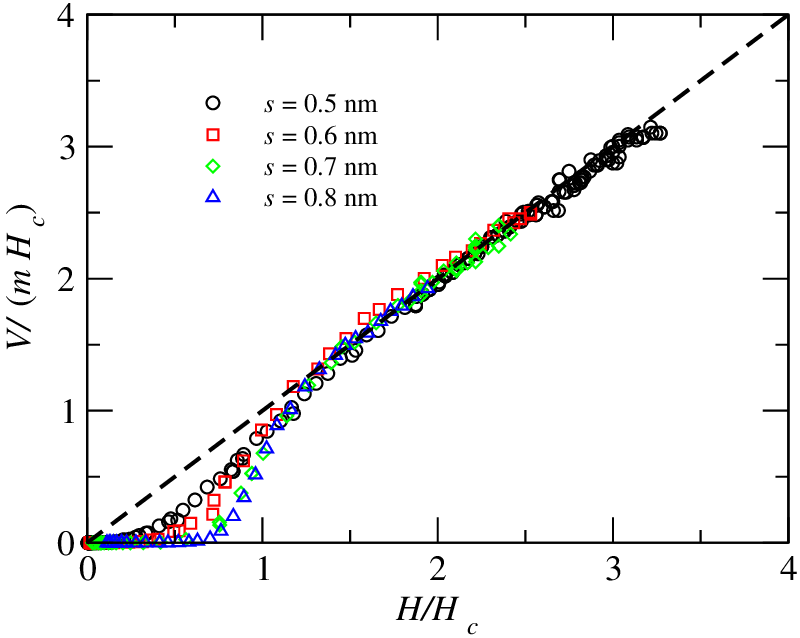}
\caption{ (color online) Velocity-field curves normalized with respect to the fast flow regime and the critical depinning filed, where the values of $H_c$ obtained in this work have been used (see Table~\ref{tab:exp}).}
\label{fig:vdef}
\end{figure}

Now that we have new estimated values for the critical depinning fields $H_c$ we can try to use scaled variables in order to obtain more information on the system. Figure~\ref{fig:vdef} shows the same data as in Fig.~\ref{fig:exp-data} but normalized with respect to the fast flow regime and with respect to the depinning field, i.e. in the scaled form $V/(m H_c)$ against $H/H_c$. One can observe in this figure that different curves are characterized by different effective temperatures. This is compatible with the fact that the disorder strength is entering not only through $H_c$ but also through the depinning temperature scale $T_c$, as expected. For increasing values of film thickness each curve seems to be at a smaller effective temperature, approaching the abrupt transition at the $T=0$ critical point. Furthermore, this is consistent with the reported values of $T_c/T$ given in Table.~\ref{tab:exp}.

\begin{figure}[!tbp]
\includegraphics[width=7.5cm,clip=true]{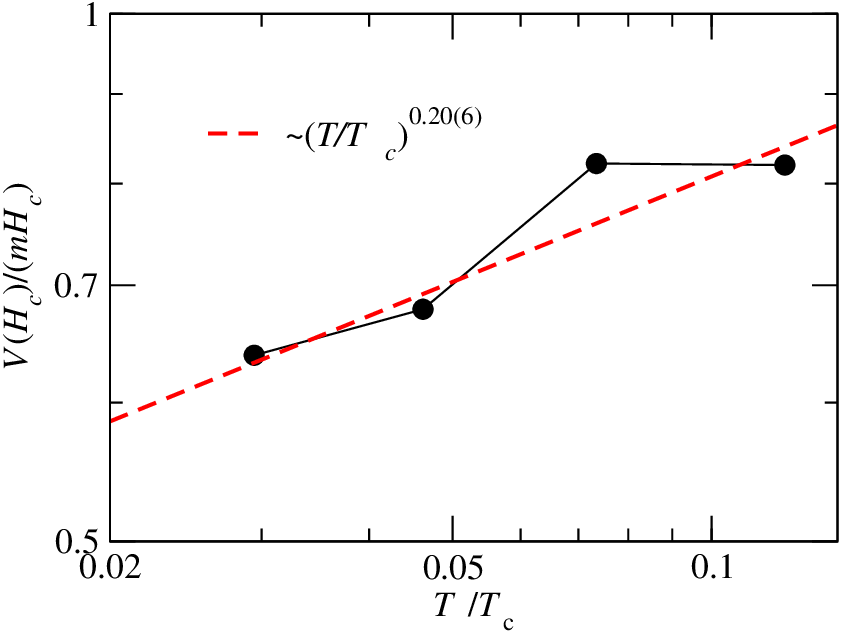}
\caption{ (color online) Plot of the normalized velocity against the effective temperature used to fit the value of the thermal rounding exponent through $V(H_c) \sim (T/T_c)^\psi$, which gives $\psi = 0.20(6)$.}
\label{fig:psi}
\end{figure}

All this information can now be used to experimentally test the predicted scaling behavior of the thermal rounding of the depinning transition. For each $V(H)$ curve in Fig.~\ref{fig:exp-data} one can now interpolate in order to obtain a velocity value corresponding to the critical field, $V(H_c)$. By plotting this value (normalized to the fast flow regime) against the effective temperature, one should be able to observe the corresponding power-law behavior defining the thermal rounding exponent, $V(H_c) \sim (T/T_c)^\psi$. Although we have only four points in the temperature scale, corresponding to the different thickness values, fitting the thermal rounding exponent from this data points gives the value $\psi = 0.20(6)$, as shown in Fig.~\ref{fig:psi}. Beyond the large error bar this is consistent with our proposed value $\psi =0.15$, which has been obtained by using numerical models for interface depinning.~\cite{bustingorry_thermal_rounding_epl,bustingorry_thermal_depinning_exponent}

\begin{figure}[!tbp]
\includegraphics[width=7.5cm,clip=true]{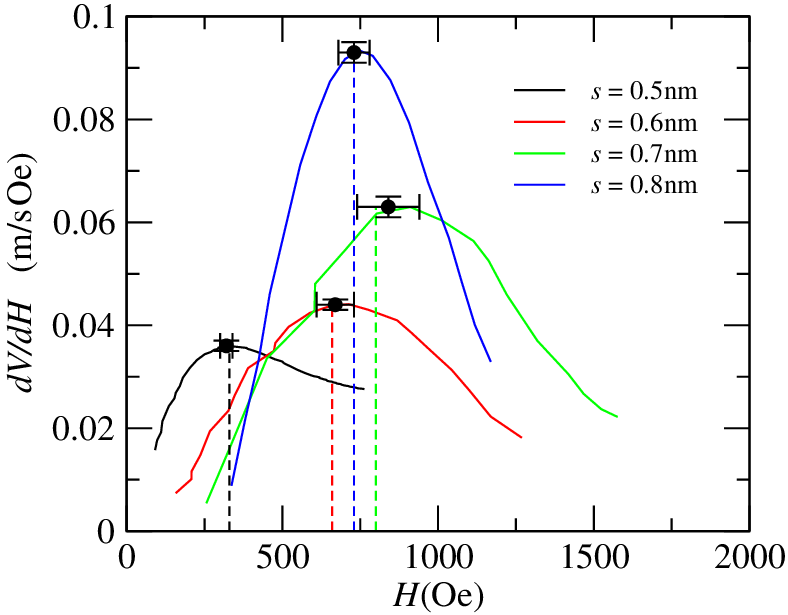}
\caption{ (color online) Phenomenological determination of the critical depinning field as the field of maximum variation of the velocity. Each line corresponds to a five order polynomial interpolation of the data in Fig.~\ref{fig:exp-data} around the depinning field. Symbols represent the obtained point in the maximum of each curve used to extract $H_c^{\mathrm{phen}}$, while the vertical dashed lines correspond to the values of $H_c$ reported in Table~\ref{tab:exp}.}
\label{fig:polynomial}
\end{figure}

We end this section showing that the values of the critical depinning field so far obtained with the protocol based on scaling arguments is consistent with a different (phenomenological) determination of the critical field. Recalling that at zero temperature the critical depinning field is the point where a maximum variation of the velocity is observed, one may wonder if this is also true at finite temperatures. In a thermally rounded velocity-field curve the inflexion point gives the field of maximum velocity variation, i.e. the the point where $dV/dH$ is maximum, which we identify with a phenomenological critical field $H_c^{\mathrm{phen}}$. In order to use this phenomenological criterion with the bare data of Fig.~\ref{fig:exp-data} we use a five order interpolation polynomial around the inflexion point, whose first order derivative is given in Fig.~\ref{fig:polynomial}. For each film thickness the maximum of the curve and the associated field value can be well characterized, as shown with symbols in Fig.~\ref{fig:polynomial}. The obtained values for $H_c^{\mathrm{phen}}$ are also included in Table~\ref{tab:exp}. We also show in the same figure, with vertical dashed lines, the values of $H_c$ previously obtained with the scaling analysis and showing a very good agreement with the phenomenological procedure. Although this phenomenological approach is not \textit{a priori} justified, the surprising agreement obtained here gives further support to the scaling theory.

\section{Universal scaling function around depinning}
\label{sec:scal}

\begin{figure}[!tbp]
\includegraphics[width=7.5cm,clip=true]{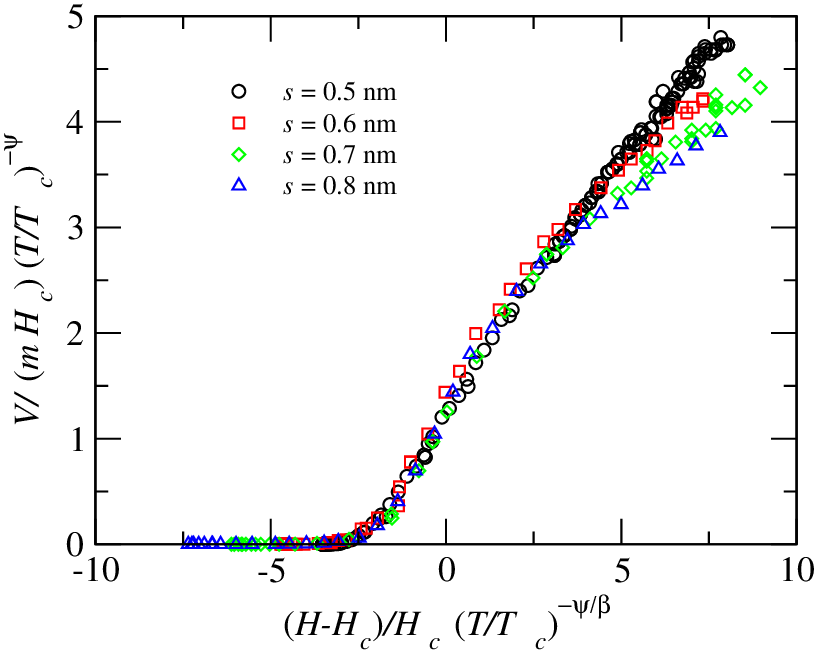}
\caption{ (color online) Universal scaling function of the velocity around the thermally rounded depinning regime, using scaling variables as in Eq.~\eqref{eq:thermal_rounding_scaling_norm}.}
\label{fig:scal}
\end{figure}

\begin{figure}[!tbp]
\includegraphics[width=7.5cm,clip=true]{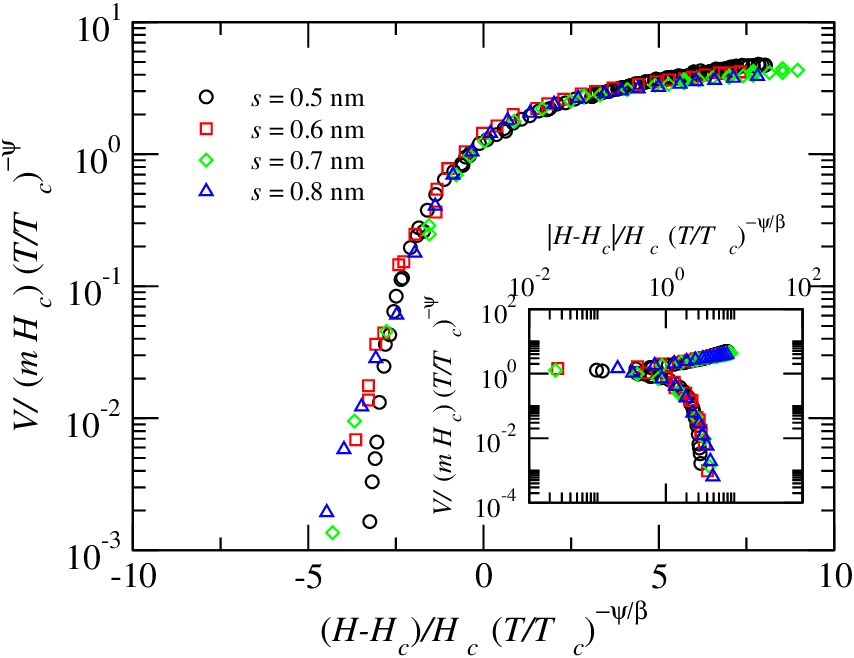}
\caption{ (color online) Universal scaling function of the velocity around the thermally rounded depinning regime. The main panel shows the same data as in Fig.~\ref{fig:scal} but in a log-linear representation which shows how the scaling fails far below the depinning threshold. For completeness, the inset shows the same data but in a log-log representation, emphasizing the good scaling close to the critical depinning region.}
\label{fig:scal-log}
\end{figure}

Now that we have obtained experimental estimates for the critical depinning field $H_c$ and the effective temperature scale $T/T_c$ for each film thickness and that we also have estimated the thermal rounding exponent $\psi$ from the experimental data, we can test the universal scaling form for the velocity within the thermal rounding regime, Eq.~\eqref{eq:thermal_rounding_scaling}. In order to do that, we gather all the information and write the scaling form using normalized quantities. Therefore, we will test in the following the universal scaling form
\begin{equation}
\label{eq:thermal_rounding_scaling_norm}
 \frac{V}{m H_c} \sim \left( \frac{T}{T_c} \right)^\psi G\left[ \frac{H-H_c}{H_c} \left( \frac{T}{T_c}
\right)^{-\psi/\beta} \right],
\end{equation}                                                                                                                                                                                                                                                                                                                                                                                                                                                                                                                        
where $G(x)$ is a universal function which is expected to behave as $G(x) \sim \mathrm{const.}$ for $x \ll 1$ and $G(x) \sim x^\beta$ for $x \sim 1$ (for $x \gg 1$ the system is outside the thermal rounding regime). We show in Fig.~\ref{fig:scal} the velocity-field curves of Fig.~\ref{fig:exp-data} in the scaled form, Eq.~\eqref{eq:thermal_rounding_scaling_norm}, using the effective temperature scale for each film thickness. The proposed scaling of the data is fairly good, making all the experimental curves in Fig.~\ref{fig:exp-data} to collapse in a single universal form within experimental tolerances. The scaling is expected to fail at large field values, as one can observe in Fig.~\ref{fig:scal} for large positive values. Furthermore, the log-linear scale used in Fig.~\ref{fig:scal-log} to represent the same data shows that the scaling works for values of the scaled variable $x= [(H-H_c)/H_c] (T/T_c)^{-\psi/\beta}$ close to zero and the scaling start failing for large negative values. Therefore, as expected, for very large positive $x$, within the fast-flow regime, or very large negative $x$, in the creep regime, the experimental data start deviating from the universal function. Finally in the inset of Fig.~\ref{fig:scal-log} we show the same scaling form but in a log-log representation (taking care of negative $x$ values) in order to emphasize the good collapse of the data for $x= [(H-H_c)/H_c] (T/T_c)^{-\psi/\beta}$ close to zero.

Finally, it is worth mentioning that Nattermann, Pokrovsky and Vinokur~\cite{nattermann_hysteresis_domainwall} proposed a phenomenological form for the full force and temperature dependence of the velocity of a domain wall in a random medium, which includes the full functional dependence. This form describes the thermal rounding regime around $H_c$ but it also includes the $H \ll H_c$ creep regime, therefore depending on the three exponents, $\beta$, $\psi$ and $\mu$. It can be shown that close to the critical depinning field the functional form proposed in Ref.~\onlinecite{nattermann_hysteresis_domainwall} reduces to the scaling form given in Eq.~\eqref{eq:thermal_rounding_scaling_norm}. In spite of this, it has been recently shown that this phenomenological functional form does not appropriately describe the numerical data for the thermally rounded depinning region.~\cite{bustingorry_thermal_depinning_exponent}

\section{Conclusions and comments}
\label{sec:conc}

The experimental data for the velocity against field obtained for the ultrathin Pt/Co/Pt films with perpendicular anisotropy give an outstanding support to the scaling ideas behind the depinning transition, particularly emphasizing the agreement with theoretical predictions for the depinning exponent $\beta$ and the thermal rounding exponent $\psi$. Each of the velocity-field curves displays the three characteristic regimes: creep, depinning and fast-flow. The creep regime, already largely accounted for in the literature, gives strong evidence for the creep exponent $\mu=1/4$. From the fast-flow regime one obtains the mobility, which gives information on the dissipation process during the flow regime and is a key parameter in order to work with normalized quantities. Besides, around the critical depinning field, our analysis unambiguously shows that the depinning exponent $\beta < 1$. Another important point is that, since the strength of the pinning disorder potential is changing with the sample thickness and the experimental data correspond to room temperature, this amounts to different effective temperatures. In fact, this effective temperature $T/T_c$ can be directly fitted from the data within the creep regime if the value of the critical depinning field is known.

We have presented in this work a scaling analysis of these experimental results based on scaling ideas behind the depinning transition and in particular of its thermal rounding. Under the key assumption that $\beta=1/3$, as suggested by many theoretical and numerical works, we have obtained improved estimates for the critical depinning field $H_c$. Besides, we have satisfactorily compared the obtained values with the phenomenological critical field estimated from the maximum variation of the velocity. This also gives strong support to the use of the value $\beta = 1/3$, characterizing the zero temperature depinning transition.

Based on our analysis one can also discard a value of the depinning exponent close to the one in the universality class of the Kardar-Parisi-Zhang (KPZ) equation with quenched disorder.~\cite{kpz,Barabasi-Stanley} The expected value would be the same as in directed percolation, $\beta_{KPZ}=\beta_{DP}=0.64$. In fact, direct integration of the quenched-KPZ equation gives $\beta_{KPZ}=0.616$.~\cite{lee2005} Here we have shown in Fig.~\ref{fig:beta-hc} that the largest values of $\beta$, obtained at the lower bound for the critical field $H^*$, are always $\beta < 0.6$ and that the scaling properties can be correctly described with $\beta =1/3$. Therefore, the experimental data can not be described within the quenched-KPZ universality class.

Furthermore, the new estimates for $H_c$ permit us to obtain an experimental estimate for the thermal rounding exponent $\psi = 0.20(6)$. This results is compatible with the numerical value $\psi=0.15$ corresponding to $(1+1)$ dimensional interface depinning within the quenched Edwards-Wilkinson universality class.~\cite{bustingorry_thermal_rounding_epl,bustingorry_thermal_depinning_exponent} Besides, this give us the opportunity to test the universal scaling form for the velocity-field curves in the thermal rounding regime. We have found that the data can be satisfactorily collapsed into a single curve with the obtained parameters.

Our results show that the experimental data can be satisfactorily described by the exponents of quenched Edwards-Wilkinson universality class. However, this equation also predicts interfaces with unbounded local relative displacements at large enough length scales, i.e. the roughness exponent characterizing the geometry of the interface is $\zeta > 1$. We therefore interpret our findings as suggesting that the velocity-field curves are effectively testing the small length scale fluctuations, given by the quenched Edwards-Wilkinson universality class. Whether the behavior at larger scales (i.e beyond the scales probed by the moving domain walls in the analyzed experiments) yields a crossover to plastic flow or to a new elastic universality class, is an interesting open question. In order to clarify this point, experiments at lower temperatures around depinning and focusing on the large scale geometrical properties, would be necessary.

\begin{acknowledgments}
The authors thank P. Metaxas, J.-P. Jamet and J. Ferr\'e for providing us the experimental data analyised in this work. T. G. acknowledge support by the Swiss National Science Fundation under MaNEP and Division II. S.B. and A.B.K. are financially supported by CONICET Grant No. PIP11220090100051.
\end{acknowledgments}

\bibliography{tfinita4}

\end{document}